\documentclass{amsart}
\usepackage[dvips]{graphicx}
\usepackage[mathscr]{eucal}
\newtheorem{Thm}{Theorem}[section]
\theoremstyle{definition}
\newtheorem{Theorem}[Thm]{Theorem}
\newtheorem{Lemma}[Thm]{Lemma}
\newtheorem{Corollary}[Thm]{Corollary}
\newtheorem{Proposition}[Thm]{Proposition}

\newtheorem{Example}[Thm]{Example}
\theoremstyle{remark}
\newtheorem{Remark}{Remark}
\font\sy=cmsy10

\font\ym=msbm10  
\newcommand{\Aut}{{\rm Aut}}

\newcommand{\R}{\text{\ym R}}

\newcommand{\C}{\text{\ym C}}

\newcommand{\cH}{{\hbox{\sy H}}}
\newcommand{\cL}{{\hbox{\sy L}}}

\newcommand{\sH}{\mathscr H}
\newcommand{\sK}{\mathscr K}

\title[]
{Geometric Mean of States and Transition Amplitudes}
\author[Yamagami Shigeru]{Yamagami Shigeru}

\begin{document}
\maketitle 
\begin{center}
Division of Mathematics and Informatics
\end{center}
\begin{center}

Ibaraki University 
\end{center}
\begin{center} 
Mito, 310-8512, JAPAN 
\end{center}    

\begin{abstract}
The transition amplitude between square roots 
of states, 
which is an analogue of Hellinger integral in classical 
measure theory, is investigated in connection with 
operator-algebraic representation theory. 
A variational expression based on geometric mean of 
positive forms is utilized to obtain an approximation 
formula for transition amplitudes. 
\end{abstract}
\bigskip

\section*{Introduction}
Geometric mean of positive operators is introduced by 
Pusz and Woronowicz in terms of associated positive 
sesquilinear forms, which is later generalized in 
various directions 
(see \cite{B,KA,LL} for example). 
When this is applied to normalized positive functionals on 
a *-algebra (so-called states), it leads us to 
an approach to the theory of positive cones 
in the modular theory of W*-algebras, 
which turns out to be closely related to 
A.~Uhlmann's transition probability 
(or fidelity) between states (see \cite{AU,U1,R}). 

We here first clarify how 
inner products
between square roots of states 
(which is referred to as ``transition amplitude'' 
in this paper just from its superficial appearance 
but without any physically serious justification)   
is relevant in representation theory of C*-algebras, 
which is then combined with Pusz-Woronowicz' geometric 
mean to get a variational expression for 
our transition amplitudes. 
The result is especially useful in establishing 
an approximation formula, which says that, 
if states $\varphi$ and $\psi$ of a C*-algebra $A$ 
are restricted to an increasing sequence of
C*-subalgebras $A_n$ with the induced states of $A_n$ 
denoted by $\varphi_n$ and $\psi_n$ respectively, then we have 
\[
(\varphi^{1/2}|\psi^{1/2}) 
= \lim_{n \to \infty} (\varphi_n^{1/2}|\psi_n^{1/2})
\]
under a mild assumption on the density of $\cup_n A_n$ in 
$A$. 

A decomposition theory of transition amplitudes is also 
described in the framework of W*-algebras for 
further applications. 

\section{Geometric Mean of Positive Forms}
We begin with reviewing Pusz-Woronowicz' geometric mean 
on positive forms in a slightly modified fashion 
from the original account. 

Let $\alpha, \beta$ be positive (sesquilinear) forms 
on a complex vector space $H$. 
By a \textbf{representation} of an unordered pair 
$\{ \alpha, \beta\}$, we shall mean 
a linear map $j: H \to K$ of $H$ into a Hilbert space $K$ 
together with 
(possibly unbounded) positive self-adjoint operators 
$A$, $B$ on $K$ such that $A$ commutes with $B$ in the strong 
sense, 
$j(H)$ is a core for the self-adjoint operator $A + B$
and 
\[
\alpha(x,y) = (j(x)|Aj(y)), 
\quad 
\beta(x,y) = (j(x)|Bj(y)) 
\]
for $x, y \in H$. 
Note that, from the core condition, 
$j(H)$ is included in the domains of 
$A = \frac{A}{A+B+I} (A+B+I)$, 
$B = \frac{B}{A+B+I} (A+B+I)$ 
($I$ being the identity operator) and therefore in 
the domains of $A^{1/2}$ and $B^{1/2}$. 
When $A$ and $B$ are bounded, we say that 
the representation is \textbf{bounded}. 
Note that, the core condition is reduced to 
the density of $j(H)$ in $K$ for a bounded representation. 

A hermitian form $\gamma$ on $H$ 
is said to be \textbf{dominated} by 
$\{ \alpha, \beta\}$ if 
$|\gamma(x,y)|^2 \leq \alpha(x,x)\,\beta(y,y)$ for $x, y \in H$. 
Note that the order of $\alpha$ and $\beta$ 
is irrelevant in the domination. 

\begin{Theorem}[Pusz-Woronowicz] 
Let $(j:H \to K, A,B)$ be a representation of 
positive forms $\alpha, \beta$ on $H$. Then, for $x \in H$,  
we have the following variational expression. 
\begin{align*}
(A^{1/2}j(x)|B^{1/2}j(x)) 
&= \sup \{ \gamma(x,x); 
\text{$\gamma$ is positive and 
dominated by $\{ \alpha, \beta\}$} \}\\
&= \sup \{ \gamma(x,x); 
\text{$\gamma$ is dominated by $\{ \alpha, \beta\}$} \}.
\end{align*}
\end{Theorem}

The positive form defined by the right hand side of 
the theorem is 
called the \textbf{geometric mean} of $\{ \alpha, \beta \}$ 
and denoted by $\sqrt{\alpha\beta} = \sqrt{\beta\alpha}$. 

\section{$L^2$-Analysis on Quasi-equivalence of States}

Associated to a W*-algebra $M$, we have 
the standard Hilbert space $L^2(M)$ so
that its positive cone consists of symbols
$\varphi^{1/2}$ where $\varphi$ varies 
in the set $M_*^+$ of normal positive linear 
functionals of $M$. On the Hilbert space $L^2(M)$, 
$M$ is represented by compatible left and right actions 
in such a way that 
\[
(\varphi^{1/2}|x\varphi^{1/2}) = \varphi(x) 
=  (\varphi^{1/2}|\varphi^{1/2}x)
\quad
\text{for $x \in M$}
\]
(inner products being linear in the second variable 
by our convention). 
This type of vectors are known to satisfy 
the following inequalities (Powers-St\o rmer-Araki) 
\[
\|\varphi^{1/2} -\psi^{1/2}\|^2 \leq \| \varphi - \psi\| 
\leq \| \varphi^{1/2} - \psi^{1/2}\|\, 
\| \varphi^{1/2} + \psi^{1/2}\|.
\]
Note that, given a central projection $q$ in $M$, 
we have the following natural identifications 
for the reduced W*-algebra $qM = Mq$: 
\[
(qM)_* = qM_* = M_*q, 
\quad 
L^2(qM) = qL^2(M) = L^2(M)q.
\]
Also note that there is a natural bilinear 
map $L^2(N)\times L^2(N) \to N_* = L^1(N)$ such that 
$\varphi^{1/2}\times \varphi^{1/2}$ is mapped to $\varphi$. 
The evaluation map $N_* \ni \varphi 
\mapsto \varphi(1) \in \C$ is 
also denoted by $\langle \varphi\rangle = \varphi(1)$ 
in this paper, which satisfies trace property 
$\langle \varphi^{1/2}\psi^{1/2} \rangle 
= \langle \psi^{1/2}\varphi^{1/2} \rangle$. 

If $\varphi$ is faithful, 
we denote by $\Delta_\varphi$ and $J_\varphi$ 
the associated modular operator and modular conjugation 
respectively.  
The positive (self-adjoint) operator 
$\Delta_\varphi^{1/2}$ has a linear subspace 
$M\varphi^{1/2}$ as a core and we see 
\[
\Delta_\varphi^{1/2}(x\varphi^{1/2}) 
= \varphi^{1/2}x
\quad 
\text{and}
\quad 
J_\varphi(x\varphi^{1/2}) = \varphi^{1/2}x^*. 
\]
More generally, if $\psi$ is another positive normal 
functional of $M$, 
then the half-powered relative modular operator 
$\Delta_{\psi,\varphi}^{1/2}$ contains $M\varphi^{1/2}$ 
as a core and we have 
$\Delta_{\psi,\varphi}(x\varphi^{1/2}) = 
\psi^{1/2}x$ for $x \in M$. 
Consult \cite{AA, MTB} for systematic accounts on 
all these operations other than the standard texts on 
modular theory such as \cite{BR1,T1,T2}. 

\begin{Lemma}
For a positive normal functional $\omega$ of a W*-algebra, 
let $e$ and $z$ be 
its support and central support respectively. 
Then we have the equalities
\begin{gather*}
L^2(zM) = zL^2(M) = L^2(M)z = \overline{M\omega^{1/2}M},\\ 
eL^2(M) = \overline{\omega^{1/2}M}, 
\quad
L^2(M)e = \overline{M\omega^{1/2}}, 
\quad 
eL^2(M)e = L^2(eMe). 
\end{gather*}
Here bar denotes the closure in $L^2(M)$.

Furthermore, 
$\overline{M\omega^{1/2}} = \overline{\omega^{1/2}M}$ 
if and only if $\omega$ is faithful on $zM = Mz$. 
\end{Lemma}

\begin{proof} 
For $x \in M$, 
\[
(\omega^{1/2}x|\omega^{1/2}e) = \omega(ex^*) = \omega(x^*) 
= (\omega^{1/2}x|\omega^{1/2})
\]
shows that $\omega^{1/2}(1-e) = 0$; 
$\overline{M\omega^{1/2}} \subset L^2(M)e$. 

Let $\pi$ be a normal representation of $M$ 
on $\overline{M\omega^{1/2}}$ given by left multiplication. 
Since the projection to the subspace 
$\overline{M\omega^{1/2}}$ commutes with the left action of 
$M$, 
we can find a projection $p$ in $M$ such that 
$\overline{M\omega^{1/2}} = L^2(M)p$. 
Particularly we have 
$\omega(1-p) = \omega^{1/2}\omega^{1/2}(1-p) = 0$ and therefore 
$e \leq p$. 

Let $Q$ be the projection to the subspace 
$\overline{M\omega^{1/2}M} \subset L^2(M)$. Then $Q$ is realized 
by multiplication of a central projection $q$ of $M$. 
From $(1-q)\omega = (1-q)\omega^{1/2}\omega^{1/2} = 0$, we see that 
$z \leq q$; $L^2(M)z \subset \overline{M\omega^{1/2}M}$. 
On the other hand, $x\omega^{1/2}yz = x\omega^{1/2}zy 
= x\omega^{1/2}y$ shows the reverse inclusion. 

Assume that $\overline{M\omega^{1/2}} = \overline{\omega^{1/2}M}$. 
If $x \in zM$ satisfies $\omega(x^*x) = 0$, i.e., 
$x\omega^{1/2} = 0$, then  
\[
xL^2(zM) = \overline{xM\omega^{1/2}M} 
= \overline{x\omega^{1/2}MM}
= 0
\]
and hence $x = 0$. 
Conversely, if $\omega$ is faithful on $zM$, 
the associated vector 
$\omega^{1/2}$ is cyclic and separating for $zM$; 
\[
\overline{M\omega^{1/2}} = L^2(zM) = \overline{\omega^{1/2}M}.
\]

Let $\omega_e$ be the restriction of $\omega$ 
to $eMe$, which is faithful. 
Since $e$ commutes with $\omega$, 
the relation $a\omega_e^{1/2} = \omega_e^{1/2}b$ with 
$a, b \in eMe$ implies 
$a\omega^{1/2} = \omega^{1/2}b$ by the reduction relation 
for modular operators 
(a consequence of Connes' $2\times 2$ matrix analysis 
and more results can be found in \cite{PT}), 
which gives the unitarity of 
\[
L^2(eMe) \ni x \omega_e^{1/2}y 
\mapsto x\omega^{1/2}y \in eL^2(M)e.
\] 
\end{proof}

\begin{Remark}~ 
The support projection $e$ is characterized 
as the minimal one among projections $p$ in $M$ satisfying 
$\overline{M\omega^{1/2}} = L^2(M)p$. 
\end{Remark}

\begin{Corollary}
Let $\varphi$ and $\psi$ be states of a C*-algebra $A$. 
\begin{enumerate}
\item 
$\varphi$ and $\psi$ are disjoint if and only if 
$A\varphi^{1/2}A$ and $A\psi^{1/2}A$ are orthogonal.
\item
$\varphi$ and $\psi$ are quasi-equivalent if and only if 
$\overline{A\varphi^{1/2}A} = \overline{A\psi^{1/2}A}$. 
\item 
The state $\varphi$ is pure if and only if 
$\overline{A\varphi^{1/2}} \cap \overline{\varphi^{1/2}A} 
= \C \varphi^{1/2}$. 
\end{enumerate}
\end{Corollary} 

\begin{proof}
Given a state $\varphi$ of a C*-algebra $A$, let $z(\varphi)$ 
be the central support of $\varphi$ in the universal envelope 
$A^{**}$. Then it is well-known 
(see \cite[Chapter~3]{P} for example) 
that $\varphi$ and $\psi$ are disjoint (resp.~quasi-equivalent) 
if and only if $z(\varphi)z(\psi) = 0$ 
(resp.~$z(\varphi) = z(\psi)$). 
Since 
$\overline{A\varphi^{1/2} A} = 
\overline{A^{**}\varphi^{1/2} A^{**}}$ in $L^2(A^{**})$, 
(i) and (ii) 
are consequences of the lemma.

Let $e$ be the support of $\varphi$ in $A^{**}$. Then the identity 
\[
\overline{A\varphi^{1/2}} \cap \overline{\varphi^{1/2}A} 
= L^2(A^{**})e \cap eL^2(A^{**}) = L^2(eA^{**}e)
\]
shows that the condition in (iii) 
is equivalent to $eA^{**}e = \C e$, 
i.e., the purity of $\varphi$. 
\end{proof}

Let $\omega$ be a state of a C*-algebra $A$ and
$\{ \tau_t \in \Aut(A) \}_{t \in \R}$ be 
a one-parameter group of *-isomorphisms. 
Recall that $\omega$ and $\{ \tau_t \}$ satisfy 
the \textbf{KMS-condition} if the following requirements 
are satisfied: 
Given $x, y \in A$, the function $\R \ni t \mapsto 
\omega(x\tau_t(y))$ is analytically extended to 
a continuous function on the strip
$\{ \zeta \in \C; -1 \leq \Im \zeta \leq 0 \}$ so that 
$\omega(x\tau_t(y))|_{t=-i} = \omega(yx)$. 

If one replaces $y$ with $\tau_s(y)$ and $x$ with $1$, then 
the condition takes the form 
$\omega(\tau_{s-i}(y)) = \omega(\tau_s(y))$ for $s \in \R$ and 
we see that the analytic function $\omega(\tau_z(y))$ 
is periodically extended to an entire analytic function. 
Thus $\omega(\tau_t(y))$ is a constant function of $t$; 
the automorphisms $\tau_t$ make $\omega$ invariant. 

\begin{Lemma}
If $\omega$ satisfies the KMS-condition, 
then $\overline{A\omega^{1/2}} = \overline{\omega^{1/2}A}$. 
\end{Lemma}

\begin{proof}
We argue as in \cite{BR2}: By the invariance of $\omega$, 
a unitary operator $u(t)$ in $\overline{A\omega^{1/2}}$ is 
defined by $u(t)(x\omega^{1/2}) = \tau_t(x)\omega^{1/2}$, which is 
continuous in $t$ from the continuity assumption on 
the function 
$\omega(x\tau_t(y))$. Moreover, the function 
$\R \ni t \mapsto u(t)x\omega^{1/2}$ is analytically continued 
to the strip $\{ -1 \leq \Im\zeta \leq 0 \}$. 
By Kaplansky's density theorem and analyticity preservation 
for local uniform convergence, 
the same property holds for $x \in A^{**}$ and 
the KMS-condition takes the form 
\[
(x\omega^{1/2}|u(t)y\omega^{1/2})|_{t=-i} 
= (\omega^{1/2}x|\omega^{1/2}y)
\quad 
\text{for $x, y \in A^{**}$.}
\]
Let $z$ be the central support of $\omega$ in $A^{**}$ and 
assume that $a \in zA^{**}$ satisfies $a\omega^{1/2} = 0$. 
Then, $xa\omega^{1/2} = 0$ for $x \in A^{**}$ 
and therefore 
$(\omega^{1/2}(xa)|\omega^{1/2}y) = 0$ for any $y \in A^{**}$ 
by analytic continuation, whence $\omega^{1/2}xa = 0$ for 
$x \in A^{**}$. Thus $zL^2(A^{**})a = 0$ and we have 
$a = 0$. 
\end{proof}

As a simple application of our analysis, we record here 
a formula which describes 
transition amplitude between purified states. 
First recall the notion of purification on states 
introduced by S.L.~Woronowicz (\cite{W}): 
Given a state $\varphi$ of a C*-algebra $A$, 
its purification $\Phi$ is a state on 
$A\otimes A^\circ$ defined by 
\[
\Phi(a\otimes b^\circ) = 
\langle \varphi^{1/2} a\varphi^{1/2}b \rangle. 
\]
Here $A^\circ$ denotes the oppositve algebra of $A$ with 
$a \mapsto a^\circ$ denoting the natural antimultiplicative 
isomorphism. 

From the above definition, 
$(a\otimes b^\circ) \Phi^{1/2} \mapsto a\varphi^{1/2} b$
gives rise to a unitary isomorphism 
$\overline{(A\otimes A^\circ)\Phi^{1/2}} 
\cong \overline{A\varphi^{1/2} A}$ and  
the GNS-representation of $A\otimes A^\circ$ 
with respect to $\Phi$ generates 
the von Neumann algebra $M\vee M'$ with 
$M = A^{**} z(\varphi)$ represented on 
$\overline{A \varphi^{1/2} A}$ by left multiplication. 
Thus $\varphi$ is a factor state 
if and only if $\Phi$ is a pure state.  
Moreover, two factor states $\varphi$ and $\psi$ of $A$
are quasi-equivalent if and only if their purifications 
are equivalent. 

\begin{Proposition}
Let $\varphi$ and $\psi$ be factor states of a C*-algebra 
$A$ with their purifications denoted by 
$\Phi$ and $\Psi$ respectively. Then we have 
\[
(\Phi^{1/2}|\Psi^{1/2}) = 
(\varphi^{1/2}|\psi^{1/2})^2.
\]
\end{Proposition}

\begin{proof}
In view of the equalities
\[
(\varphi^{1/2}|\psi^{1/2}) = 0 = (\Phi^{1/2}|\Psi^{1/2}).
\] 
for disjoint $\varphi$ and $\psi$, 
we need to consider the case that $\varphi$ and $\psi$ are 
quasi-equivalent, i.e., $z(\varphi) = z(\psi)$.
Since $\varphi$ and $\psi$ are assumed to be factor states, 
their purifications $\Phi$ and $\Psi$ are pure with 
the associated GNS-representations of 
$A\otimes A^\circ$ generate the full operator algebra 
$\cL(L^2(M))$. Thus, through the obvious identification 
$L^2(\cL(L^2(M))) = L^2(M)\otimes L^2(M)$, 
$\Phi^{1/2}$ and $\Psi^{1/2}$ correspond to 
$\varphi^{1/2}\otimes \varphi^{1/2}$ and 
$\psi^{1/2}\otimes \psi^{1/2}$ respectively, whence we have 
\[
(\Phi^{1/2}|\Psi^{1/2}) 
= (\varphi^{1/2}\otimes \varphi^{1/2} | 
\psi^{1/2}\otimes \psi^{1/2}) 
= (\varphi^{1/2}|\psi^{1/2})^2.
\]
\end{proof}

\begin{Remark}
For purifications of states on a commutative C*-algebra, 
we have 
$(\Phi^{1/2}|\Psi^{1/2}) 
= (\varphi^{1/2} | \psi^{1/2})$. 
The general case is a mixture of these two formulas. 
\end{Remark}

\begin{Lemma}
Let $\pi: A \to M$ be a homomorphism from a C*-algebra $A$ 
into a W*-algebra $M$ 
and assume that $\pi(A)$ is *-weakly 
dense in $M$. 
Then we have an isometry $T: L^2(M) \to L^2(A^{**})$ 
such that $T(\pi(a)\varphi^{1/2}\pi(b)) = 
a(\varphi\circ \pi)^{1/2}b$ 
if $\varphi \in M_*^+$ and $a, b \in A$. 
\end{Lemma}

\begin{proof}
Since the map $M_* \ni \varphi \mapsto \varphi\circ\pi \in A^*$ is 
norm-continuous (in fact it is contractive), 
we have $A^{**} \to M$ as its transposed map, which 
is $\pi$ when restricted to $A \subset A^{**}$. 
In other words, we see that $\pi$ is extended to 
a normal homomorphism $\widetilde\pi: A^{**} \to M$ 
of W*-algebras in such a way that, if
$\varphi\circ\pi$ is regarded as a normal functional on $A^{**}$, 
it is equal to $\varphi\circ\widetilde\pi$. 

By our weak*-density assumption, 
$\widetilde\pi$ is surjective and 
and we can find a central projection $z \in A^{**}$ so that 
$\ker\widetilde\pi = zA^{**}$ and 
$(1 - z)A^{**} \cong M$ by $\widetilde\pi$. From the relation 
\[
z(\varphi\circ\pi) = z(\varphi\circ \widetilde\pi) = 
\widetilde\pi(z)(\varphi\circ \widetilde\pi) = 0,
\] 
one sees that the isomorphism 
$M_* \cong zA^*$ takes the form $\varphi \mapsto 
(1-z)(\varphi\circ\pi) = \varphi\circ\pi$, which yields 
the formula in question by taking square roots. 
\end{proof}

\begin{Corollary}
Let $\pi:A \to B$ be a *-homomorphism between C*-algebras and 
$\varphi$, $\psi$ be positive functionals of $B$. 
Assume that, given $a \in A$ and $b \in B$, we can find a norm-bounded sequence 
$\{ a_n\}_{n \geq 1}$ in $A$ such that 
\[
\lim_{n \to \infty} 
\pi(a_n)\pi(a)\varphi^{1/2} = b\pi(a)\varphi^{1/2}, 
\quad  
\lim_{n \to \infty} 
\pi(a_n)\pi(a)\psi^{1/2} = b\pi(a)\psi^{1/2}. 
\]
Then we have
\[
(\varphi^{1/2}|\psi^{1/2}) = ((\varphi\circ\pi)^{1/2}|(\psi\circ\pi)^{1/2}).
\]
\end{Corollary}

\begin{proof}
Let $z(\varphi)$ and $z(\psi)$ be the central projections in $B^{**}$ specified by 
\[
\overline{B\varphi^{1/2}B} = z(\varphi)L^2(B^{**}), 
\quad 
\overline{B\psi^{1/2}B} = z(\psi) L^2(B^{**}). 
\]
Let $M = (z(\varphi)\vee z(\psi)) B^{**}$ and 
$\pi_M: A \to M$ be a homomorphism 
defined by $\pi_M(a) = (z(\varphi)\vee z(\psi))\pi(a)$. 

Let $\rho$ be the direct sum of GNS-representations associated to $\varphi$ and $\psi$. 
Then $\rho$ is supported by $z(\varphi)\vee z(\psi)$: $\rho$ is extended to an isomorphism 
of $(z(\varphi)\vee z(\psi))B^{**}$ onto $\rho(B)''$. 
On the other hand, by the approximation assumption, $\rho(\pi(A))$ is dense in $\rho(B)$ 
with respect to the strong operator topology. 
Thus $\pi_M(A)$ is *-weakly dense in $M$ and 
the lemma can be applied if one notices 
that $(z(\varphi)\vee z(\psi))\varphi = \varphi$ and $(z(\varphi)\vee z(\psi))\psi = \psi$ 
as identities in the predual of $B^{**}$. 
\end{proof}

\begin{Example}
Consider quasifree states $\varphi_S$ and $\varphi_T$ of a CCR C*-algebra $C^*(V,\sigma)$. 
Let $(\ |\ )$ be a positive inner product in $V$ majorizing both of $S+\overline S$ and $T+\overline T$. 
For example, one may take 
$(x|y) = (S+ \overline{S} + T + \overline{T})(x,y)$ 
as before. 
Then the presymplectic form $\sigma$ is 
continuous relative to $(\ |\ )$ and, if we let $V'$ be the associated 
Hilbert space (i.e., the completion of $V/\ker(\ |\ )$ with respect to $(\ |\ )$), 
$\sigma$ induces a presymplectic form $\sigma'$ on $V'$. 
Moreover, $S$ and $T$ also give rise to polarizations 
$S'$ and $T'$ on the presymplectic vector space 
$(V', \sigma')$ respectively. 

Let $\pi: C^*(V,\sigma) \to C^*(V',\sigma')$ be the *-homomorphism induced from 
the canonical map $V \to V'$ ($\pi(e^{iv}) = e^{iv'}$ 
if $v'$ represents the quotient of $v$). 
Then $\pi$ satisfies the approximation condition with respect to 
quasifree states associated to $S'$ and $T'$ 
(see the proof of Proposition~4.3). 
Since $\varphi_S = \varphi_{S'}\circ \pi$ 
and similarly for $T$, we obtain 
\[
(\varphi_S^{1/2}| \varphi_T^{1/2}) 
= (\varphi_{S'}^{1/2}| \varphi_{T'}^{1/2}).
\] 
Thus local positions of square roots of 
quasifree states are described under the assumption that 
$V$ is complete and $\sigma$ is continuous with respect to a non-degenerate inner product. 
\end{Example}

\section{Transition Amplitude between States}

Let $\omega$ be a positive functional of a C*-algebra $A$. 
According to \cite{PW},  
we introduce two positive sesquilinear forms $\omega_L$ 
and $\omega_R$ on $A$ defined by 
\[
\omega_L(x,y) = \omega(x^*y), 
\qquad 
\omega_R(x,y) = \omega(yx^*), 
\quad 
x, y \in A.
\]

\begin{Lemma}
Let $M$ be a W*-algebra and 
let $\varphi$, $\psi$ be positive normal functionals of $M$. 
Then 
\[
\sqrt{\varphi_L\psi_R}(x,y) 
= \langle \varphi^{1/2} x^*\psi^{1/2}y \rangle
\quad
\text{for $x, y \in M$.}
\]
\end{Lemma}

\begin{proof}
By the positivity
$\langle \varphi^{1/2} x^*\psi^{1/2}x \rangle 
= (x\varphi^{1/2}x^*|\psi^{1/2}) \geq 0$ 
and the Schwarz inequality 
$|\langle \varphi^{1/2}x^*\psi^{1/2}y\rangle|^2 
\leq 
\varphi(x^*x) \psi(yy^*)$, the positive form 
$(x,y) \mapsto \langle \varphi^{1/2} x^*\psi^{1/2}y \rangle$ 
is dominated by $\{ \varphi_L, \psi_R \}$.  

Assume for the moment that $\varphi$ and $\psi$ are faithful
and consider the embedding 
$j: M \ni x \mapsto x\varphi^{1/2} \in L^2(M)$. Then 
$\varphi_L$ is represented by the identity operator, 
whereas 
$\psi(xx^*) = \| \psi^{1/2}x\|^2$ 
shows that $\psi_R$ is represented by the relative modular 
operator $\Delta$ with 
$\Delta^{1/2}(x\varphi^{1/2}) = \psi^{1/2} x$. 
Recall that $M\varphi^{1/2}$ is a core for $\Delta^{1/2}$. 
Thus Theorem~1.1 gives 
\[
\sqrt{\varphi_L\psi_R}(x,y) 
= (x\varphi^{1/2}| \Delta^{1/2}(y\varphi^{1/2})) 
= (x\varphi^{1/2}|\psi^{1/2}y) = 
\langle \varphi^{1/2}x^*\psi^{1/2}y \rangle.
\] 

Now we relax $\varphi$ and $\psi$ to 
be not necessarily faithful. 
Let $e$ be the support projection of 
$\varphi+\psi$. Then it is the supprot for 
$\varphi_n = \varphi + \frac{1}{n}\psi$ 
and $\psi_n = \frac{1}{n}\varphi + \psi$ as well. 
In particular, $\varphi_n$ and $\psi_n$ are faithful 
on the reduced algebra $eMe$. 

Let $\gamma$ be a positive form on $M$ dominated by 
$\{ (\varphi_n)_L, (\psi_n)_R \}$. Then 
$\varphi_n(1-e) = 0 = \psi_n(1-e)$ shows that 
\[
|\gamma(x(1-e), (1-e)y)|^2 
\leq \varphi_n((1-e)x^*x(1-e)) \psi_n((1-e)yy^*(1-e)) = 0, 
\]
i.e., $\gamma(x,y) = \gamma(xe,ey)$ for $x, y \in M$, whence 
we have 
\[
\gamma(x,y) = \gamma(xe,ey) = 
\overline{\gamma(ey,xe)} = 
\overline{\gamma(eye,exe)} 
= \gamma(exe,eye).
\]
Since the restriction $\gamma|_{eMe}$ is dominated 
by $(\varphi_n|_{eMe})_L$ and $(\psi_n|_{eMe})_R$ with 
$\varphi_n$ and $\psi_n$ faithful on $eMe$, we have 
\[
\gamma(x,x) = \gamma(exe,exe) 
\leq \langle e\varphi_n^{1/2}ex^*e \psi_n^{1/2} ex\rangle 
= \langle \varphi_n^{1/2}x^*\psi_n^{1/2}x \rangle. 
\] 
Taking the limit $n \to \infty$, we obtain 
$\gamma(x,x) \leq \langle \varphi^{1/2}x^*\psi^{1/2}x \rangle$ 
in view of the Powers-St\o rmer inequality. 
\end{proof}

\begin{Remark}~ 
\begin{enumerate}
\item
The case $\varphi = \psi$ was dealt with in 
the proof of \cite[Theorem~3.1]{PW} 
under the separability assumption on $M_*$.
\item
In the notation of \cite{U2}, 
we have 
$QF_t(\varphi_L,\psi_R)(x,y) = \langle 
\varphi^{1-t}x^*\psi^ty \rangle$
for $0 \leq t \leq 1$ and $x, y \in M$. 
\end{enumerate} 
\end{Remark}

Given a positive functional $\varphi$ of a C*-algebra $A$, 
let $\widetilde\varphi$ be the associated normal functional 
on the W*-envelope $A^{**}$ through 
the canonical duality pairing. 

\begin{Lemma}
Let $\varphi$ and $\psi$ be positive functionals 
on a C*-algebra $A$ 
with $\widetilde\varphi$ and $\widetilde\psi$ the corresponding 
normal functionals on $A^{**}$. Then 
\[
\sqrt{\varphi_L\psi_R}(x,y) = 
\langle {\widetilde \varphi}^{1/2} x^* {\widetilde\psi}^{1/2}y 
\rangle 
\quad
\text{for $x, y \in A \subset A^{**}$.}
\]
\end{Lemma}

\begin{proof}
The positive form $A\times A \ni (x,y) \mapsto 
\langle {\widetilde \varphi}^{1/2} x^* {\widetilde\psi}^{1/2}y 
\rangle$ (recall that $x^*{\widetilde\psi}^{1/2}x$ 
is in the positive cone to see the positivity) is dominated by 
${\widetilde\varphi}_L$ and ${\widetilde\psi}_R$ because of 
\[
|\langle {\widetilde \varphi}^{1/2} x^* {\widetilde\psi}^{1/2}y 
\rangle|^2 
\leq {\widetilde\varphi}(x^*x) {\widetilde\psi}(yy^*) 
= \varphi(x^*x) \psi(yy^*).
\]
Consequently, 
\[
\langle {\widetilde \varphi}^{1/2} x^* {\widetilde\psi}^{1/2}x 
\rangle 
\leq \sqrt{\varphi_L\psi_R}(x,x) 
\quad 
\text{for $x \in A$.}
\] 

To get the reverse inequality, let $\gamma$ be a positive form on 
$A\times A$ dominated by $\varphi_L$ and $\psi_R$. Then we have the domination 
inequality 
\[
|\gamma(x,y)|^2 \leq 
\varphi(x^*x) \psi(yy^*) 
= \| x{\widetilde\varphi}^{1/2}\|^2\, 
\| {\widetilde\psi}^{1/2}y\|^2. 
\]
Since $A$ is dense in $A^{**}$ relative to the $\sigma^*$-topology, 
we see that $\gamma$ is extended to 
a positive form $\widetilde\gamma$ on $A^{**}\times A^{**}$ so that 
\[
|\widetilde\gamma(x,y)|^2 
\leq \| x{\widetilde\varphi}^{1/2}\|^2\, 
\| {\widetilde\psi}^{1/2} y\|^2
\quad 
\text{for $x, y \in A^{**}$,}
\]
whence 
\[
\gamma(x,x) = \widetilde\gamma(x,x) \leq 
\sqrt{{\widetilde\varphi}_L {\widetilde\psi}_R}(x,x) 
= \langle {\widetilde \varphi}^{1/2} x^* {\widetilde\psi}^{1/2}x 
\rangle 
\quad 
\text{for $x \in A$.}
\]
Maximization on $\gamma$ then yields the inequality 
\[
\sqrt{\varphi_L\psi_R}(x,x) \leq 
\langle {\widetilde \varphi}^{1/2} x^* {\widetilde\psi}^{1/2}x 
\rangle 
\quad 
\text{for $x \in A$}
\]
and we are done. 
\end{proof}

\begin{Corollary}
Given a normal state $\varphi$ of a W*-algebra $M$, let 
$\widetilde\varphi$ be the associated normal state of the second dual 
W*-algebra $M^{**}$. Then 
\[
L^2(M) \ni \varphi^{1/2} \mapsto {\widetilde\varphi}^{1/2} 
\in L^2(M^{**})
\]
defines an isometry of $M$-$M$ bimodules. 
\end{Corollary}

\begin{proof}
Combining two lemmas just proved, we have 
\[
\langle \varphi^{1/2} x^*\psi^{1/2}y \rangle 
= 
\sqrt{\varphi_L\psi_R}(x,y) 
= 
\langle {\widetilde \varphi}^{1/2} x^* {\widetilde\psi}^{1/2}y 
\rangle 
\]
for $x, y \in M$. 
\end{proof}

In what follows, $\varphi^{1/2}$ is identified with 
${\widetilde\varphi}^{1/2}$ via the isometry just established: 
Given a positive normal functional $\varphi$ of 
a W*-algebra $M$, 
$\varphi^{1/2}$ is used to stand for 
a vector commonly contained 
in the increasing sequence of Hilbert spaces 
\[
L^2(M) \subset L^2(M^{**}) \subset L^2(M^{****}) \subset \dots.
\]

In accordance with this convention, the formula in the previous 
lemma is simply expressed by 
\[
(x\varphi^{1/2}|\psi^{1/2}y) = \sqrt{\varphi_L\psi_R}(x,y) 
\quad
\text{for $x, y \in A$.}
\]
Here the left hand side is the inner product in $L^2(A^{**})$, 
whereas 
the right hand side is the geometric mean of positive forms on 
the C*-algebra $A$. 
Note that, the formula is compatible with 
the invariance of geometric means: 
$\sqrt{\varphi_L\psi_R}(x,y) = \sqrt{\psi_L\varphi_R}(y^*,x^*) 
= \sqrt{\varphi_R\psi_L}(y^*,x^*)$.

\begin{Remark}~ 
\begin{enumerate}
\item 
When $\varphi$ and $\psi$ are vector states of 
a full operator algebra $\cL(\sH)$ associated to 
normalized vectors $\xi, \eta$ in $\sH$, the inner product 
$(\varphi^{1/2}|\psi^{1/2})$ is reduced to 
the transition \textit{probability} $|(\xi|\eta)|^2$. 
Moreover, in view of the inequality 
$t \varphi^{1/2} + (1-t)\psi^{1/2} 
\leq (t\varphi + (1-t)\psi)^{1/2}$ for $0 \leq t \leq1$
(which follows from 
$(\varphi^{1/2} - \psi^{1/2})^2 \geq 0$), 
our transition amplitude meets the requirements for 
transition probability listed in \cite{S}
\item
Let $P(\varphi,\psi)$ be the transition probability between 
states in the sense of A.~Uhlmann (\cite{U1}). 
Then we have 
$P(\varphi,\psi) = 
\langle |\varphi^{1/2}\psi^{1/2}| \rangle^2$ 
(cf.~\cite{R}) and 
\[
(\varphi^{1/2}|\psi^{1/2})^2 
\leq P(\varphi,\psi) \leq 
(\varphi^{1/2}|\psi^{1/2}). 
\]
\end{enumerate}
\end{Remark}

\section{Approximation on Transition Amplitudes}

In this section, we shall see how transition amplitudes 
are approximated by states obtained by restriction 
to subalgebras. 

\begin{Lemma}[cf.~{\cite[Proposition~17]{U2}}]
Let $\Phi: A \to B$ be a unital Schwarz map between unital 
C*-algebras. Then, for positive linear functionals 
$\varphi, \psi$ of $B$, 
\[
(\varphi^{1/2}|\psi^{1/2}) \leq 
((\varphi\circ\Phi)^{1/2}| (\psi\circ \Phi)^{1/2}). 
\]
\end{Lemma}

\begin{proof}
Let $\gamma: B \times B \to \C$ be a positive form 
dominated by $\{ \varphi_L,\psi_R\}$. 
Then 
\[
|\gamma(\Phi(x),\Phi(y))|^2 \leq \varphi(\Phi(x)^*\Phi(x)) 
\psi(\Phi(y)\Phi(y)^*) 
\leq 
\varphi(\Phi(x^*x)) \psi(\Phi(yy^*))
\]
shows that the positive form 
$A \times A \ni (x,y) \mapsto \gamma(\Phi(x),\Phi(y))$ is 
dominated by 
$\{ (\varphi\circ \Phi)_L, (\psi\circ \Phi)_R \}$. 
Thus 
\[
\gamma(1,1) = \gamma(\Phi(1),\Phi(1)) 
\leq \sqrt{(\varphi\circ\Phi)_L(\psi\circ\Phi)_R}(1,1) 
= ((\varphi\circ\Phi)^{1/2} | (\psi\circ\Phi)^{1/2}). 
\]
Maximizing $\gamma(1,1)$ with respect to $\gamma$, we obtain the 
inequality. 
\end{proof}

\begin{Theorem}
Let $\varphi$ and $\psi$ be positive functionals on 
a C*-algebra $A$ with unit $1_A$. 
Let $\{ A_n \}_{n \geq 1}$ be 
an increasing sequence of C*-subalgebras of $A$ containing  
$1_A$ in common and assume 
that, given any $a \in A$, 
we can find a sequence $\{ a_n \in A_n \}_{n \geq 1}$ 
satisfying 
\[
\lim_{n \to \infty} a_n\varphi^{1/2} = a\varphi^{1/2}, 
\quad 
\lim_{n \to \infty} \psi^{1/2}a_n = \psi^{1/2}a
\]
in norm topology. 
Set $\varphi_n = \varphi|_{A_n}, 
\psi_n = \psi|_{A_n} \in A_n^*$. Then 
the sequence $\{ (\varphi_n^{1/2}|\psi_n^{1/2}) \}_{n \geq 1}$ 
is decreasing and converges to $(\varphi^{1/2}|\psi^{1/2})$. 
\end{Theorem}

\begin{proof}
The sequence $\{ (\varphi_n^{1/2}|\psi_n^{1/2}) \}$ is decreasing 
with $(\varphi^{1/2}|\psi^{1/2})$ a lower bound 
by the previous lemma. 

Let $e_n$ and $f_n$ be projections on $L^2(A^{**})$ defined by 
\[
e_n L^2(A^{**}) = \overline{A_n \varphi^{1/2}}, 
\quad 
f_nL^2(A^{**}) = \overline{\psi^{1/2}A_n}. 
\]
Choose positive forms 
$\gamma_n: A_n \times A_n \to \C$ for $n \geq 1$ so that 
$\gamma_n$ is dominated by $\{ (\varphi_n)_L, (\psi_n)_R \}$ and 
satisfies 
\[\gamma_n(1,1) \geq (\varphi_n^{1/2}|\psi_n^{1/2}) - 1/n. 
\]
From the domination estimate on $\gamma_n$, we can find 
a linear map $C_n': \overline{\psi^{1/2} A_n} 
\to \overline{A_n \varphi^{1/2}}$ such that 
\[
\gamma_n(x,y) = (x\varphi^{1/2}|C_n'(\psi^{1/2}y))
\quad 
\text{for $x, y \in A_n$,}
\]
which satisfies $\| C_n'\| \leq 1$. 
Let $C_n = e_nC_n'f_n: \overline{\psi^{1/2}A} \to 
\overline{A\varphi^{1/2}}$. 
Since $\| C_n \| \leq 1$, we may assume that $C_n \to C$ in 
weak operator topology 
by passing to a subsequence if necessary. Now set 
\[
\gamma(x,y) = (x\varphi^{1/2}|C(\psi^{1/2}y)), 
\]
which is a sesquilinear form on $A$ satisfying 
$|\gamma(x,y)| \leq \| x\varphi^{1/2}\|\, \| \psi^{1/2}y\|$. 
Moreover, if $x \in A_m$ for some $m \geq 1$, 
\[
\gamma(x,x) = \lim_{n \to \infty} (x\varphi^{1/2}| C_n(\psi^{1/2}x)) 
= \lim_{n \to \infty} \gamma_n(x,x) \geq 0
\]
shows that $\gamma$ is positive on 
$\displaystyle \bigcup_{m \geq 1} A_m$ and hence on $A$ 
by the approximation assumption. 
Thus, $\gamma$ is a positive form dominated by 
$\{ \varphi_L, \psi_R \}$ and we have 
\begin{align*}
(\varphi^{1/2}|\psi^{1/2}) 
&\geq \gamma(1,1) = \lim_{n \to \infty} (\varphi^{1/2}|C_n\psi^{1/2})
= \lim_{n \to \infty} \gamma_n(1,1)\\ 
&\geq \lim_{n \to \infty} 
\left(
(\varphi_n^{1/2}|\psi_n^{1/2}) - \frac{1}{n} 
\right) = \lim_{n \to \infty} (\varphi_n^{1/2}|\psi_n^{1/2}).
\end{align*}
\end{proof}

As a concrete example, 
we have the following situation in mind: 
Let $(V,\sigma)$ be a real presymplectic vector space and 
$C^*(V,\sigma)$ be the associated C*-algebra. 
Let $\varphi$ and $\psi$ be quasifree states of $C^*(V,\sigma)$ 
asscoated to covariance forms $S$ and $T$ respectively: 
\[
\varphi(e^{ix}) = e^{-S(x,x)/2}, 
\quad 
\psi(e^{ix}) = e^{-T(x,x)/2}
\quad 
\text{for $x \in V$.}
\]
Note that $S$ is a positive form on the complexification 
$V^\C$ satisfying 
$S(x,y) - \overline{S(x,y)} = i\sigma(x,y)$ for 
$x, y \in V$ and similarly for $T$. 

Let $\{ V_n \}_{n \geq 1}$ be an increasing sequence of 
subspaces of $V$ and assume that 
$\displaystyle \bigcup_{n \geq 1}V_n$ is dense in $V$ with respect 
to the inner product 
\[
V\times V \ni (x,y) \mapsto (x|y) \equiv 
S(x,y) + \overline{S(x,y)} 
+ T(x,y) + \overline{T(x,y)}
\in \R.
\]
Note that $(x|x)$ may vanish on non-zero $x \in V$. 
Let $A_n$ be the C*-sualgebra of $A = C^*(V,\sigma)$ generated by 
$\{ e^{ix}; x \in V_n \}$. 

\begin{Proposition}
In the setting described above, the increasing sequence 
$\{ A_n \}_{n \geq 1}$ satisfies the approximation property 
with respect to $\varphi$, $\psi$. 
\end{Proposition}

\begin{proof}
For $x \in V$, choose a sequence $x_n \in V_n$ so that 
$(x_n - x| x_n - x) \to 0$. Then, for any $y \in V$, 
\[
\| (e^{ix_n} - e^{ix}) e^{iy}\varphi^{1/2}\|^2
= 2 - 2e^{-S(x_n-x)/2} \Re e^{i\sigma(x_n-x,y+x/2)}
\to 0
\]
beacuse of the continuity of $\sigma$ 
with respect to $(\ |\ )$. 
Similarly, we see 
$\| \psi^{1/2}e^{iy}(e^{ix_n} - e^{ix}) \|^2 \to 0$. 
Since any $a \in C^*(V,\sigma)$ is approximated in norm 
by a finite linear combination of $e^{ix}$'s, we are done. 
\end{proof}

\section{Central Decomposition}

In this final section, we describe 
a decomposition theory 
for transition amplitudes between normal states, 
which will be effectively used in \cite{GQFS}. 

Let $M$ be a W*-algebra with a separable predual and $Z$ be 
a central W*-subalgebra of $M$. Since 
$Z$ is a commutative W*-algebra, we have an expression 
$Z = L^\infty(\Omega)$ with $\Omega$ 
a measurable space furnished with a measure class $d\omega$. 
If we choose a measure $\mu$ 
representing the measure class $d\omega$, 
then it further induces a decomposition of the form
\[
\int_{L^\infty(\Omega)}^\oplus M_\omega\, d\omega
\quad 
\text{on}
\quad
L^2(M) = \int_{L^2(\Omega)}^\oplus 
L^2(M_\omega)\, \mu(d\omega). 
\]
Here $\{ M_\omega \}$ is a measurable family of W*-algebras 
and each normal functional $\varphi$ of $M$ is expressed by 
a measurable family 
$\{ \varphi_\omega \}_{\omega \in \Omega}$ 
of normal functional in such a way that   
\[
\varphi(x) = \int_\Omega 
\varphi_\omega(x_\omega)\, \mu(d\omega)
\quad
\text{if}
\ 
x = \int_{L^\infty(\Omega)}^\oplus 
x_\omega\, d\omega
\]
and the $L^2$-identification is given by 
\[
(\varphi^{1/2}|\psi^{1/2}) 
= \int_\Omega 
(\varphi_\omega^{1/2}|\psi_\omega^{1/2})\, 
\mu(d\omega).
\]

This can be seen as follows: 

\begin{Lemma}
Let $\{ M_\omega, \cH_\omega\}$ be a measurable family 
of von Neumann algebras and set 
\[
M = \int_{L^\infty}^\oplus M_\omega\,d\omega, 
\quad 
\cH = \int_{L^2}^\oplus \cH_\omega\,\mu(d\omega).
\]
Let $\xi = \int^\oplus \xi_\omega \mu(d\omega)$ 
be a vector in $\cH$. 
Then $\xi$ is cyclic for $M$ if and only if 
$\xi_\omega$ is a cyclic 
vector of $M_\omega$ for a.e.~$\omega$.
\end{Lemma}

\begin{proof}
The `only if' part follows from the fact that 
\[
\int_{L^2}^\oplus (M_\omega\xi_\omega)^\perp\, 
\mu(d\omega)
\]
is a subspace orthogonal to $M\xi$. 

For the `if' part, we use the commutant formula 
\[
M' = \int_{L^\infty}^\oplus M_\omega'\,d\omega
\]
and check that, if $\xi_\omega$ is a separating 
vector of $M_\omega'$ for a.e.~$\omega$, then 
\[
x' \xi = \int_{L^2}^\oplus x'_\omega \xi_\omega\, 
\mu(d\omega) = 0
\]
for $x' = \int^\oplus x'_\omega\,d\omega \in M'$ implies 
$x'_\omega \xi_\omega = 0$ for a.e.~$\omega$ and 
hence $x'_\omega = 0$ for a.e.~$\omega$, i.e., $x' = 0$. 
\end{proof}

By replacing $(M_\omega,\sH_\omega)$ with 
$(M_\omega\otimes 1_\sK,\sH_\omega\otimes \sK)$
($\sK$ being a Hilbert space) and then restricting 
to cyclic subspaces, we may assume that 
we can find a cyclic and separating $\xi$ for the von Neumann 
algebra $M$. 
Then $\xi_\omega \in \sH_\omega$ 
is a cyclic and separating vector 
of $M_\omega$ for a.e.~$\omega$. 
Let $J_\omega$ be the modular conjugation associated to 
$\xi_\omega$ (these are defined up to null sets). 
Then, from the relevant definitions on modular stuff, 
we see that 
$\{J_\omega\}$ is a measurable family of operators and 
\[
J = \int_{L^\infty}^\oplus J_\omega\,d\omega
\]
gives the modular conjugation associated to $\xi$. 

Let 
\[
\varphi = \int_{L^1(\Omega)}^\oplus 
\varphi_\omega\,\mu(d\omega), 
\quad
\psi = \int_{L^1(\Omega)}^\oplus \psi_\omega\,\mu(d\omega)
\]
and choose $\xi$ so that both of $\varphi$ and $\psi$ 
is majorized by the functional $(\xi|\cdot\xi)$. 
By a Radon-Nikodym type theorem (cf.~\cite{A}), 
we can find $a, b \in M$ such that 
$\varphi$, $\psi$ are associated to the vectors 
$a\xi a^* = aJaJ\xi$, $b\xi b^* = bJbJ\xi$ respectively. 
(In the notation of \cite{AA}, we can choose 
$a = \varphi^{1/4}\xi^{-1/2}$, $b = \psi^{1/4}\xi^{-1/2}$.) 
Then 
$\varphi_\omega$ and $\psi_\omega$ are represented by 
vectors $a_\omega J_\omega a_\omega J_\omega \xi_\omega$ 
and 
$b_\omega J_\omega b_\omega J_\omega \xi_\omega$ 
for a.e.~$\omega$. 
Thus, for $x, y \in M$, 
\[
\langle x_\omega \varphi_\omega^{1/2} 
y_\omega \psi_\omega^{1/2}\rangle
= 
(a_\omega \xi_\omega a_\omega^*x_\omega^*| 
y_\omega b_\omega \xi_\omega b_\omega^*)
\] 
is a measurable function of $\omega$ and we have 
\[
\langle x\varphi^{1/2}y\psi^{1/2}\rangle 
= \int_\Omega \
\langle x_\omega \varphi_\omega^{1/2} 
y_\omega \psi_\omega^{1/2}\rangle\, \mu(d\omega).
\]

Consequently, $\{ L^2(M_\omega) \}$ is a measurable family 
of Hilbert spaces 
in such a way that for any $\varphi \in M_*^+$, 
the decomposed components $\{ \varphi_\omega^{1/2} \}$ 
is measurable. Moreover, we have a decomposable unitary 
\[
L^2(M) \ni x\phi^{1/2} \mapsto 
\int_{L^2(\Omega)}^\oplus x_\omega \phi_\omega^{1/2} y_\omega.
\]

We shall now rewrite the results so far to fit into 
representation theory of C*-algebras. 

Let $\{ A(\omega) \}$ be 
a family of quotient C*-algebras of 
a C*-algebra $A$ indexed by elements in a standard Borel 
space $\Omega$. Denote by $a(\omega) \in A(\omega)$ 
the quotient element of $a \in A$. 

A positive functional $\varphi$ of a C*-algebra $A$ is said to 
be \textbf{separable} if the Hilbert space 
$\overline{A\varphi^{1/2}A}$ is separable. 

A family 
$\{ \varphi_\omega \in A(\omega)^*_+ 
\}_{\omega \in \Omega}$ of positive functionals 
is said to be \textbf{measurable} if $\varphi_\omega$ 
is separable for each $\omega \in \Omega$ and 
\[
\omega \mapsto \langle 
a(\omega)\varphi_\omega^{1/2} b(\omega)
\varphi_\omega^{1/2}\rangle 
\]
is measurable for $a, b \in A$. 
Given a measurable family of states 
$\{ \varphi_\omega \}$, 
$\{ \overline{A(\omega)\varphi_\omega^{1/2} A(\omega)} 
\}_{\omega \in \Omega}$ is a measurable family of 
Hilbert spaces in an obvious way. 
If we are further given a probability measure 
$\mu$, 
\[
\varphi(x) = \int_\Omega \varphi_\omega(x(\omega))\,
\mu(d\omega) 
\]
defines a state of $A$. 

A measurable 
family of positive functionals $\{ \varphi_\omega \}$ 
is said to be \textbf{disjoint} with respect to 
a probability measure $\mu$ of $\Omega$ 
if $\int_{\Omega'} \varphi_\omega \mu(d\omega)$ 
and $\int_{\Omega''} \varphi_{\omega} \mu(d\omega)$ 
are disjoint for $\Omega' \cap \Omega'' = \emptyset$. 

\begin{Proposition}~ 
\begin{enumerate}
\item 
Given a probability measure $\mu$ and 
a $\mu$-disjoint family of separable 
states $\{ \varphi_\omega \}$, 
the integrated state $\varphi = \int \varphi_\omega 
\mu(d\omega)$ is separable and we have a unitary map 
\[
\overline{A\varphi^{1/2}A} 
\ni a \varphi^{1/2} b \mapsto 
\int_{L^2(\Omega)}^\oplus 
a(\omega)\varphi_\omega^{1/2} b(\omega) 
\mu(d\omega) 
\in \int_{L^2(\Omega)}^\oplus 
\overline{A(\omega) \varphi_\omega^{1/2} A(\omega)}\, 
\mu(d\omega).
\]
\item 
Let $\{ \psi_\omega\}$ be another $\mu$-disjoint  
family of separable states 
with $\psi = \int \psi_\omega \mu(d\omega)$ 
the integrated state of $A$. 
Then, for $a, b \in A$, 
$\langle a(\omega)\varphi^{1/2} b(\omega) \psi^{1/2} \rangle$ 
is measurable and 
\[
\langle a \varphi^{1/2} b \psi^{1/2}\rangle 
= \int_\Omega 
\langle a(\omega) \varphi_\omega^{1/2} b(\omega) 
\psi_\omega^{1/2}\rangle 
\mu(d\omega).
\]
\end{enumerate}
\end{Proposition}

\begin{proof}
Let $M$ be the von Neumann algebras generated by 
the left multiplication of $A$ on 
$\overline{A\varphi^{1/2}A}$. Then, by the disjointness 
of $\{ \varphi_\omega\}$, $L^\infty(\Omega,\mu) \subset M$ 
and we can apply the results on W*-algebras. 
\end{proof}

\end{document}